# Flexibility of nucleic acids: from DNA to RNA

Bao Lei (鲍磊), Zhang Xi (张曦), Jin Lei (金雷), and Tan Zhi-Jie (谭志杰)[*]

*Department of Physics and Key Laboratory of Artificial Micro- and Nano-structures of Ministry of Education, School of Physics and Technology, Wuhan University, Wuhan 430072, China*

## Abstract

The structural flexibility of nucleic acids plays a key role in many fundamental life processes, such as gene replication and expression, DNA-protein recognition, and gene regulation. To obtain a thorough understanding of nucleic acid flexibility, extensive studies have been performed using various experimental methods and theoretical models. In this review, we will introduce the progress that has been made in understanding the flexibility of nucleic acids including DNAs and RNAs, and will emphasize the experimental findings and the effects of salt, temperature, and sequence. Finally, we will discuss the major unanswered questions in understanding the flexibility of nucleic acids.

PACS: 87.10.Pq, 87.14.G-, 87.15.ap, 87.15.H-, 87.15.La

Key words: DNA, RNA, structure, flexibility, salt, temperature, sequence, computer simulation

___________________________________________________________

[*]Corresponding author: zjtan@whu.edu.cn

# 1. Introduction

Nucleic acids are negatively charged biopolymers and their structures are generally stabilized by base pairing/stacking interactions and metal ion-binding.[1−4] Due to the polymeric nature of nucleic acids and the stabilizing energy on the order of $\sim k_B T$ (thermal energy), nucleic acids generally show strong conformational fluctuations and are rather flexible. The flexibility of nucleic acids is extremely important for their biological functions such as gene replication and expression, protein recognition, and gene regulation.[1−5]

To evaluate the flexibility of nucleic acids, various experimental methods have been developed, such as atomic force microscopy (AFM),[6,7] fluorescence resonance energy transfer (FRET),[8,9] small-angle x-ray scattering (SAXS),[9−11] magnetic tweezers (MT),[12−14] and optical tweezers (OT),[14−16] among others. Some theoretical models have also been developed and combined with experimental approaches to quantify the flexibility of nucleic acids, such as the worm-like chain (WLC) model and free-joint chain model.[17,18] Recently, with the development of computational methods, molecular dynamics (MD) simulations[19−24] have been widely used to examine the flexibility of nucleic acids at the atomic level. Beyond the experiments that can only evaluate the macroscopic properties of flexibility, MD simulations and theoretical modeling at the atomic level can reveal detailed microscopic information, such as conformational changes of nucleic acids and ion binding patterns, as well as related microscopic mechanisms.[21,24] The use of various advanced experimental techniques and atomistic MD/modeling has greatly enhanced the understanding of nucleic acid flexibility.[6−22]

Due to the order of $\sim k_B T$ for base pairing/stacking energy and the polyanionic nature of nucleic acids,[1−5,25] their flexibilities strongly depend on their sequences, salt ions in solution, and temperature, which would affect the strength of base pairing/stacking, ion binding, and chain conformational entropy, respectively. Drugs or proteins can also interact with nucleic acids and dramatically affect their structures and flexibilities.[2,3,5,26] In addition, nucleic acids can exhibit distinctively different flexibilities for different structural states depending on the temperature and ionic conditions.[27−31] Therefore, nucleic acid flexibility is influenced by several critical factors, such as sequence, salt, and temperature.

In this review, we will focus on recent progress made in understanding the flexibility of nucleic acids. Since the family of nucleic acids includes single-stranded (ss) DNA/RNA, double-stranded (ds) DNA/RNA, and a large number of RNA tertiary folds, the main text is organized as follows. First, we will provide a brief overview of the flexibility of ssDNA/RNA. Second, we will focus on the flexibility of dsDNA, which has already attracted much attention for many years. Third, we will describe recent progress in understanding the flexibility of RNAs. Finally, we will discuss the major unanswered questions in understanding the flexibility of nucleic acids.

## 2. Flexibility of ssDNA and ssRNA

The ss chain is an elementary structural and functional segment of nucleic acids. For example, RNA structures generally consist of different types of loops, and the ss chain also represents the denatured state of nucleic acids.[1,32,33] Furthermore, the ss chain is an important intermediate in many key biochemical processes, such as replication, recombination repair, and transcription, and is specifically recognized by many proteins.[32,34] The flexibility of the ss chain plays a significant role in its interactions with other macromolecules, such as proteins.[35] Generally, under the physiological conditions, ss nucleic acid chains composed of generic sequences are rather flexible, and can be approximately described using the free-joint chain model, while ss nucleic acid flexibility may be sensitive to the sequence and ionic environment.[36−52]

The flexibilities of ss nucleic acids have been quantified using various experimental approaches, such as force-extension curves,[36−40] FRET,[41,42] SAXS,[43] AFM,[44] and fluorescence recovery after photobleaching.[52] These experimental measurements are summarized in Table 1. In addition, computer simulations, such as atomistic molecular and Monte Carlo simulations, have been employed to evaluate the flexibility of ss nucleic acids, including the effects of sequence and salt.[36−52]

### 2.1. Sequence effect

Various experiments have suggested that the structure and flexibility of an ss DNA/RNA chain strongly depends on the intra-chain interactions, such as base-pairing and base stacking, which are highly correlated with the nucleic acid sequence. An ssDNA chain composed of a generic sequence can fold into a secondary structure, such as a hairpin or a helix through base-pairing/stacking and different sequence compositions give generic ss chains different properties, including flexibility.[36,40,45] Single-molecule experiments have suggested that an ssDNA can be modeled as the free-joint-chain model under physiological ionic conditions with a small persistence length of ~7.5 Å,[49] since strong intra-chain base stacking is generally not observed in these molecules.

A homo-polynucleotide ss DNA may exhibit self-stacking interactions between nearby nucleotide bases, which can be sufficiently strong to maintain the rigidity of the ss chain.[44] The varying strengths of intra-chain self-stacking interactions results in different flexibilities of the four types of homo-polynucleotide ss DNA molecules, including poly(dA), poly(dT), poly(dC), and poly(dG). Previous experiments have shown that poly(dT) is much more flexible than poly(dC), poly(dG), and poly(dA), since the self-stacking interactions in these molecules were much stronger than that in poly(dT). Actually, poly(dT) behaves as a free polyelectrolyte chain,[39,44,50] while poly(A), poly(G), and poly(C) would form ordered ss helices because of strong intra-chain self-stacking.[33,39]

Since thymine (T) and uridine (U) show similar strengths in base-stacking interactions, the effect of sequence on the flexibility of ssRNA is very similar to that of DNA.[39,51] A previous study showed that the intra-chain self-stacking of ssRNA is slightly stronger than that of ssDNA because of the difference in the sugar ring. Slightly lower flexibility for poly(rU) than for poly(dT) has been observed under the same ionic conditions.[47]

However, the details of self-stacking interactions in ssDNA/RNA and the mechanism of the flexibility of ssDNA/RNA remain unclear. Advanced experimental techniques and improved force fields for atomistic simulations can be used to examine these issues.

## 2.2. Effects of salt and chain length

Due to the similar charge density on ssDNA and ssRNA, the ion dependence of ss RNA flexibility is very similar to that of ssDNA.[41,47,53]

Ions in solution can bind to ssDNA/RNA, which would increase nucleic acid flexibility by neutralizing the negative charges on the phosphates. Numerous experiments have shown that ssDNA/RNA becomes more flexible with increasing concentrations of ions, including $Na^+$ and $Mg^{2+}$, and this ion-dependence of flexibility is stronger in longer sequences.[36,38,41–43,47,48] $Mg^{2+}$ has a higher ionic charge than $Na^+$, and experiments and simulations have shown that $Mg^{2+}$ is approximately 60−120-fold more efficient than $Na^+$ in ionic neutralization.[41,47] Additionally, the ion concentration dependence of ss DNA/RNA flexibility is stronger for $Na^+$ than for $Mg^{2+}$.[36,38,41,47] At high salt concentration, intra-sequence self-stacking interactions would dominate the global flexibility.[42,47] For ss generic sequences, empirical formulas have been derived for ion-dependent persistence length, which are practically useful.[43,47,48] For an ssDNA under force, a previously derived formula can describe the force-dependent ssDNA force-extension curves at various concentrations of NaCl.[54]

Previous studies have also shown that the ion-dependence of ssDNA/RNA flexibility depends on sequence length, i.e., the persistence length $P$ of longer ssDNA/RNAs increases more rapidly than that of short sequences when the ion concentration is decreased.[36,38,41,47,53] This occurs because ion binding to ssDNA/RNAs strongly depends on sequence length.[36,38,41−43,47,52] Experiments have also shown that there is scaling law between the size of ssDNA and its length, and the scaling exponent decreases with increasing monovalent salt concentration.[55]

However, the effect of multivalent ions on ssDNA/RNA flexibility remains unclear since the strong ion-ion correlation and possible intra-chain base-pairing/stacking can become tightly coupled.

## 3. Flexibility of dsDNA

### 3.1. General features in flexibility of dsDNA

Since the discovery of the dsDNA helix, numerous studies have demonstrated that dsDNAs are extensively involved in various life processes.[1,2,56,57] For example, dsDNA can fold into compact structures to enter into bacteriophage heads[56] or form chromatin,[57] and dsDNA can also sharply bend on a local scale to execute its biological functions, such as replication, DNA repair, and transcription, among other functions.[1,2] Therefore, understanding the biological processes related to dsDNAs requires the comprehensive understanding of dsDNA flexibility.

DsDNA, as a highly dynamic structure, is stretchable, bendable, and twistable *in vivo* and *in vitro*, and its flexibility can be characterized by the three important elastic parameters: stretching modulus $S$, bending persistence length $P$, and torsional persistence length $C$. $S$, $P$, and $C$ describe the stretching, bending, and twisting flexibilities, respectively. In addition, contour length ($L$), end-to-end distance ($R_{ee}$), and radius of gyration ($R_g$) have also been used to describe the global structural flexibility of dsDNAs. Extensive experiments have been conducted to quantify the flexibility of dsDNA, and the experimental measurements are summarized in Table 2.

Previous reviews[1,2] have described the bending persistence length $P$ of long dsDNA in buffers of moderate salt concentration, which was 45−50 nm based on early experiments. Recently, advanced single-molecule techniques have enabled accurate measurements of dsDNA flexibility. Herrero-Galán et al. manipulated long dsDNAs using magnetic tweezers and optical tweezers, and they observed a bending persistence length $P$ of ~49 ± 2 nm and a stretch modulus $S$ of ~935 ± 121 pN at 150 mM NaCl.[15] Other recent force-extension experiments for dsDNA, indicated that torsional persistence length $C$ was 100 ± 7 nm,[58] a slightly higher value than is generally accepted.[59−62] Very recently, Dekker et al. explored the elastic properties of dsDNA and derived all four elastic constants for dsDNA at 100 mM monovalent salt: 45 ± 2 nm for $P$, 1000 ± 200 pN for $S$, 109 ± 4 nm for $C$, and a negative twist-stretch coupling parameter of ~17 ± 5.[14]

Since dsDNA is generally stabilized by specific base-pairing/stacking interactions and the binding of metal ions, the flexibility of dsDNA is strongly dependent on salt, sequence, and temperature. Next, we will comprehensively summarize the flexibility of dsDNAs in the three aspects. Finally, we will introduce recent findings for short dsDNA and dsDNA under high force.

**3.2. Salt effect**

Due to the highly negative charges on dsDNA, the flexibility and stability of dsDNA are tightly coupled to the metal ions present in solution.[1,28,63,64] Numerous experimental and theoretical studies have focused on the role of salt in dsDNA flexibility and have revealed the following major features:[15,27,28,65−74]

1. The increase of monovalent salt concentration enhances the flexibility of dsDNA, which is reflected by the decrease in bending persistence length $P$. $P$ can decrease to ~45−50 nm at high (~1 M) salt concentration.[15,27,28]

2. The addition of multivalent (≥ 2+) salt clearly enhances the flexibility of dsDNA[63] and causes the collapse of long dsDNAs into compact condensates, which is reflected by a persistence length (~20−40 nm) much lower than ~50 nm.[27,28,74]

3. The stretching modulus $S$ is also strongly dependent on salt. Increasing salt concentration increases $S$, and $S$ is larger for higher valent salts.[15,28]

4. With increasing monovalent salt concentration, the contour length and twist angle both decrease linearly as a logarithmic function of salt concentration.[75,76]

However, on the salt effect on dsDNA flexibility, there are still important questions remained. First, the strength of the salt dependence of $P$ of dsDNA, particularly above ~0.1 M monovalent salt, is unclear. Odijk, Skolnick and Fixman have previously proposed the OSF theory, which divides bending persistence length into intrinsic and electrostatic contributions.[77,78] OSF theory predicts that the electrostatic part only contributes less than ~10% to the total persistence length of dsDNA under physiological ionic conditions. Nevertheless, Manning developed a model based on his counterion condensation theory by defining a hypothetical structure of dsDNA in the absence of DNA residual charges as a "null isomer of DNA."[79] Manning's model predicted that the dependence of $P$ on salt concentration is nearly log-[salt] over the entire concentration range and the electrostatic contribution to total bending persistence length can reach ~86%, which has been supported by several experiments.[67−69] Savelyev et al. developed a two-bead coarse-grained structural model for dsDNA and conducted a reexamination by MD simulations. They found that both electrostatic and nonelectrostatic interactions play comparable roles in dsDNA flexibility.[80] They also found that dsDNA bending persistence length decreases by ~25% when monovalent salt is decreased from 0.1 M to 1 M.[81]

Therefore, there is still no consensus regarding how DNA flexibility depends on salt concentration, particularly at high salt concentrations of > 1 M.

Second, how multivalent ions influence the flexibility of dsDNA is still unclear since multivalent ions interact with DNA much more strongly than monovalent ions.[70,71] As indicated in experiments, multivalent ions such as $Mg^{2+}$ and $Co(NH_3)_6^{3+}$, have very different effects on DNA flexibility at the same ionic strength compared with monovalent ions. For example, $Co(NH_3)_6^{3+}$ can lead to a value of $P$ for dsDNA as low as 25−30 nm.[28] Is the effect of multivalent ions coupled to ion-mediated effective interaction between segments in dsDNA?[82,83] Furthermore, what are the roles of divalent ions in the flexibility of dsDNA since some divalent ions can induce effective attractions between dsDNA helices, while other divalent ions cannot?[84]

In addition to the unsolved issues described above, additional questions remain at the microscopic level: (1) Why does the stretching modulus increase with increasing ion concentration and ion valence? (2) Why are stretching and twisting negatively coupled?

### 3.3. Temperature effect

For a polymer, the WLC model describes the relationship between bending rigidity $g$ and bending persistence length $P$ as:[85]

$$P = \frac{g}{k_B T}, \qquad (1)$$

where $k_B$ is the Boltzmann constant and $T$ is absolute temperature. If $g$ does not change with temperature, the temperature-dependence of $P$ can be described by Eq. (1), and $P$ decreases linearly with increasing temperature. However, dsDNA is composed of sequential base pairs unlike an ideal polymer. Since base pairing/stacking strength is on the order of ~$k_B T$, temperature plays an important role in dsDNA flexibility. With increasing temperature, bases and backbone of dsDNA fluctuate more strongly, which may cause local "melted" bubbles. If temperature becomes sufficiently high, dsDNA strands can become completely separated and exhibit a "melted" state of ss chain. Therefore, the actual temperature dependence of $P$ should be much stronger than that predicted from Eq. (1).

Approximately 40 years ago, Gray and Hearst measured the sedimentation coefficient of DNA at infinite dilution and obtained the temperature dependence of $P$.[29] However, their data showed that the temperature-dependence of $P$ is even weaker than the value predicted from Eq. (1), which assumes the temperature-independent bending rigidity $g$. Recently, dsDNA cyclization experiments have indicated that local melting in dsDNA can enhance the flexibility of dsDNA more significantly than that predicted from the WLC model in the temperature range of 23−42°C.[30] A significant decrease in apparent bending persistence length may result from potential excitations of flexible defects,[86,87] which can be excited by high temperature.[88] In order to more accurately measure the temperature dependence of persistence length, Geggier and Vologodskii employed two different approaches to determine the persistence lengths of dsDNA at different temperatures, and the data from the two independent approaches were highly consistent. The experiments showed that the bending persistence length decreased nearly linearly from 53 nm to 44 nm as temperature decreased from 5°C to 42°C, but decreased very sharply at higher temperatures, reaching ~36 nm at 60°C.[31] Corresponding theoretical modeling based on the Peyrard-Bishop-Dauxois model has also shown good agreement with the experimental data. Therefore, both of the experiments by Vologodskii et al. and the Peyrard-Bishop-Dauxois-based model showed the discrepancies with the WLC model.[89,90] Very recent experiments with tethered particle motion by Driessen et al. confirmed the above findings, and also showed that the increase of temperature can lead to a more compact structure of dsDNA and the temperature-dependent $P$ is tightly coupled to the content of GC base pairs.[91]

It is understandable that a dsDNA is more flexible at higher temperature because the ds helix is generally stabilized by base pairing/stacking interactions. However, a previous experiment showed that dsDNA flexibility can become weaker with increasing temperature in [2°C, 20°C], which was proposed to be attributed to the sequence-direct DNA curvature.[92] Such temperature-weakened flexibility for some dsDNA requires further analysis.

### 3.4. Sequence effect

It is well known that dsDNA flexibility depends on its sequence, which directly affects dsDNA stability.[1−3,93,94] Numerous previous studies have shown that different sequence arrangements can greatly influence the stability of dsDNA under bending and its ability to form kinks, which can induce base pair slide to form non-native contacts.[1−3] Simultaneously, the sequence can also influence the mechanical properties of a dsDNA in contact with proteins such as histones.[1−3]

Olson et al. have performed the statistical analysis on X-ray crystal structures and found that different sequences produce distinct flexibility, where the AA·TT step belongs to the rigid class while GG·CC and GC·GC dinucleotides are even more flexible.[93] However, this conclusion may be invalid because of the choice of system where the central regions were rich in AT content while the GC steps were segregated at the terminals.[95] Furthermore, based on statistical analysis of X-ray crystal structures of protein-oligonucleotide complexes, Olson et al.

found that the average twisting of base pair steps increases in the same order within three standard chemical classes: pyrimidine–purine, purine–purine, and purine–pyrimidine.[94] Ortiz and Pablo proposed a coarse-grained model for the effect of sequence on the overall stability and flexibility of dsDNA under bending constraints.[96] They found that longer repeated segments such as AAAAAAAA were more likely to form a kink, while short repetitive segments such as CCC were less likely to form a kink. This is because a base in the AA...AA strands can slide more easily to form non-native contacts with neighboring complementary bases in the repeated TT...TT sequence.[96]

In addition to the above theoretical approaches, scanning force microscopy and AFM have been widely employed to characterize the flexibility of DNA.[95,97,98] Scipioni et al. described the intrinsic curvature of DNAs based on scanning force microscopy images, confirming that A·T-rich sequences are more flexible than G·C-rich sequences.[95] To obtain a more detailed understanding of the sequence-dependent flexibility of dsDNA, Geggier and Vologodskii determined the bending persistence length more accurately based on a cyclization method of short DNA fragments, and their data showed that variations in $P$ induced by different sequences might be sufficiently large to affect their biological activity. Such variations can also affect the binding affinity of DNA-protein complexes, in which dsDNA segment shows sharp bending.[99] Moreover, extensive theoretical and experimental analyses showed that DNA fragments containing A-tracts exhibited a tighter bend at the 3′ end than at the 5′ end, where the "A-tract" refers to the duplex $(dA)_n \cdot (dT)_n$ and is equivalent to the "T-tract".[3,96,100] TA and AG·CT steps show higher roll angle values compared to GC and GG·CC steps, and can also bend DNA even in the absence of the A-tract.[100] In addition to the specific sequence described above, the sequence GGGCCC also showed a net bend.[101]

With the rapid development of computational facility, molecular simulation has become an important tool for exploring the effect of sequence on the flexibility of dsDNA at the atomic level.[22,102] Based on an MD study of dsDNA in the gas phase, Xiao and Liang found that all pyrimidine rings were highly flexible in either isolated or paired states, whereas the imidazole rings were relatively more rigid.[22] Lavery et al. also performed systematic MD simulations to study the nearest-neighbor effects on base pairing.[102] Their simulations suggested that to predict the sequence dependence of DNA structure and dynamics, next-nearest-neighbor interactions should be taken into account because the effect is significant.[102]

### 3.5. DsDNA under high force

As described above, the elastic properties of dsDNA, such as stretching modulus and bending persistence length can be investigated through single-molecule force-extension experiments by fitting extension-force curves to the (extensible) WLC model.[49,54,103−109] Under low stretching force (e.g., < 20 pN), dsDNA generally maintains the B-form, and the measured properties reflect the intrinsic elasticity of B-DNA with specific sequences under specific environmental conditions. However, under high stretching force (> 65 pN, near physiological ionic conditions), apparent overstretching transitions can occur in the dsDNA structure.[104−108] Such overstretching transitions are based on the early single-molecule experiments of dsDNA and have been proposed to involve a stretching-induced melting transition from dsDNA to ssDNA.[28,49,110−112] However, very recent single-molecule experiments have revealed that much more complex overstretching transitions from dsDNA to other overstretched structures occur, such as peeled ssDNA with one strand peeled from another, DNA bubble with two strands separated internally, and S-DNA with elongated base pairs.[105−108] These experimental findings were also suggested using early theoretical models.[112] These recent experiments have yielded the following major findings:[54,104−108]

1. The overstretching transitions for dsDNA under high force strongly depend on the DNA sequence (GC-rich or AT-rich), the state of dsDNA ends (end-open or end-closed), ionic condition, and temperature.[54,104−108]

2. For end-closed dsDNA, whose ends are DNA hairpins, a high stretching force induces the transition from B-DNA to S-DNA at high ionic strength, while induces the transition from B-DNA to a DNA bubble at low ionic strength. Additionally, the overstretching force for the transitions appears to be higher for CG-rich DNA and lower for AT-rich DNA.[54,108]

3. For end-open dsDNA, whose ends are without constraints, a high force induces the overstretching transition from B-DNA to S-DNA at high ionic strength, while induces the transition from B-DNA to peeled ssDNA at low ionic strength. Additionally, the overstretching force for the transition from B-DNA to peeled ssDNA is higher for GC-rich dsDNA and lower for AT-rich dsDNA, while there is no transition between S-DNA and B-DNA for AT-rich dsDNA over the wide range of ionic strength of 1 mM to 100 mM.[108]

4. The S-DNA is a new form of dsDNA with elongated base pairs, which has been continuously stretched up to approximately 70% beyond its canonical B-form contour length and is more flexible ($P$ ~10 nm) than B-DNA ($P$ ~50 nm). The DNA bubble with two internal ss strands also has much higher flexibility than B-DNA.[54]

Extensive single-molecule experiments have shown the overstretching transition is directly coupled to the stability of a dsDNA, which strongly depends on the content of GC base pairs, ionic conditions, and end-constraints. Thus, it is expected that at lower temperature and in multivalent ion solutions, a dsDNA can be stretched to an S-DNA at lower ionic strength because of the higher stability. The atomic structures of overstretched dsDNA can be modeled by all-atom MD simulations.[24]

### 3.6. Flexibility of short DNA

Numerous recent experiments have suggested that short DNAs have high flexibility compared with those in kilo-base pairs.[113−117] Cyclization experiments have shown that short dsDNAs of ~100 base pairs (bps) formed circles much faster than predicted by the WLC model mainly because of the unusually large local bend angle induced by kinking.[113,114] Another series of experiments using FRET and SAXS by Yuan et al. also suggested the higher flexibility of short DNAs of 15−89 bps, which were beyond the description of the conventional WLC model.[9] In addition, recent SAXS experiments of short DNAs of ≤ 35 bps with two end gold nanocrystals by Mathew-Fenn et al. suggested that short DNAs are at least one order of magnitude more extensible than long DNAs of kilo-bps revealed by previous single-molecule stretching experiments.[118] This high flexibility for short dsDNA has been proposed to be attributed to defect excitation, which may reduce the local bending energy of dsDNA through local DNA melting of a few base pairs, local DNA kinking, and excitation of a few base pairs of S-DNA.[86,112,115,119,120] Since the local stability of dsDNA depends strongly on sequence, ionic strength, and temperature,[2,30,31,63,64,93,94] such defect excitation may be sensitive to temperature and ionic strength.[112]

However, a similar SAXS experiment showed that the flexibility of short DNAs of 42−94 bps with two linked gold nanocrystals could be described by the WLC model with a persistence length of ~50 nm.[11] On the illusive controversy, the atomic MDs have also been employed to probe the flexibility of short DNAs of 5−50 bps, and the results showed that shorter DNA may have higher apparent flexibility, which is attributed to the higher flexibility of ~6 bps at each end.[88] Nevertheless, how to explain the experiments with labeling nanocrystals developed by Mathew-Fenn et al. and Yuan et al. at the atomic level is still required, which would assist not only the understanding of the experimental findings, but also the understanding of the effect of labeling nanocrystals on the flexibility of short biomolecules.

## 4. Flexibility of RNA

### 4.1. Flexibility of dsRNA

Recently, dsRNA has been highly valued because its role in the life cycle of a cell is now more crucial than previously considered.[7,14,15,121−123] In addition to being a central role in RNA interference,[121] dsRNAs may have the potential applications in nanomedicine and nanomaterial.[122,123] Because of these important applications, the flexibility of dsRNA has been studied in various single-molecule experiments,[7,14,15] and the experimental measurements are tabulated in Table 2.

Unlike dsDNA, which is generally in the B-form, dsRNA forms a thicker right-handed duplex in the "A-form",[124] and this special helical structure gives dsRNA different flexibilities. In an early experiment of transient electric birefringence (TEB), Hagerman et al. obtained a bending persistence length of ~60 ± 10 nm for dsRNA,[125,126] which was 20−30% larger than the accepted value for dsDNA. Therefore, dsRNA is somewhat stiffer than dsDNA. Recently, Dekker et al. obtained a mean bending persistence length of 63.8 ± 0.7 nm through force-extension measurements with magnetic tweezers and of 62 ± 2 nm using AFM measurements for long dsRNAs.[7] They also obtained a torsional persistence length of ~99 ± 5 nm at an external stretching force F = 6.5 pN, which was similar to the value for dsDNA, as well as a stretching modulus of 350 ± 100 pN, which was three-fold lower than that of dsDNA. Interestingly, they observed that dsRNA exhibits positive twist-stretch coupling, which is in contrast to dsDNA with negative twist–stretch coupling.[14] Herrero-Galán et al. systematically measured the mechanical properties of dsRNA under different ion conditions at the single-molecule level.[15] They found that the values for bending persistence length $P$ for dsRNA were consistently larger than those for dsDNA under the same ionic conditions, and that $P$ decreased with increasing salt concentration over the range of 0−500 mM NaCl, which is similar to dsDNA. Furthermore, the OT measurements revealed that the stretching modulus $S$ of dsRNA increased with increasing salt concentration, in agreement with the trend for dsDNA,[28] while $S$ for dsRNA was much lower than that of dsDNA under the same salt conditions.[16,28]

Therefore, compared with dsDNA, dsRNA has a larger bending persistence length and a similar torsional persistence length. However, the stretching modulus of dsRNA is nearly three-fold lower than that of dsDNA, and a surprising difference between dsRNA and dsDNA is that dsRNA has a positive twist-stretch coupling parameter while dsDNA has a negative one. The microscopic mechanism for the apparent difference in the flexibility between dsRNA and dsDNA remains unclear and requires further investigation.

### 4.2. Flexibility of structured RNA

Generally, RNAs fold into more complex native structures rather than keep a denatured ss chain or a perfect duplex.[127,128] The flexibility of complex RNA structures is important for their biological functions, such as RNA-protein recognition and gene regulation.[129,130] Understanding the flexibility of structural RNAs would enable the deep exploration of their biological functions and related applications such as structure-based drug design.[131,132] However, there have been few extensive studies on examining the flexibility of structural RNAs. In Table 3, we summarized the existing experimental measurements for the elastic properties of structured RNAs beyond the states of the ss chain and helix.[133−143]

For RNA hairpins with a bulge loop (and an internal loop), Zacharias and Hagerman performed a series of TEB experiments.[134,137,138] For RNA hairpins with a bulge loop, the bending angles at the junction induced by bulge loops of different sizes and base compositions were determined, and were found to increase monotonically with the increment from ~8° to ~20° for both bulge loops of $A_n$ and $U_n$ as n was increased from 1 to 6 in the absence of $Mg^{2+}$. Moreover, for bulge loops constituted by $U_n$, the presence of $Mg^{2+}$ reduced the increment of bending angle by a factor of 2 for all of n values.[137] However, for RNA hairpins with symmetric internal loops with the forms $A_n$-$A_n$ and $U_n$-$U_n$ (n = 2, 4, 6), it was found that the internal loops could only distort RNAs by values that were much smaller than their bulge loop counterparts.[138]

Other experimental methods have also been employed to unravel the flexibility of structured RNAs. Al-Hashimi et al. performed a series studies on the HIV-1 TAR RNA with NMR and residual dipolar coupling (RDC), in combination with coarse-grained modeling by Brooks et al.[139,144] Because of the high resolution of the method, detailed dynamic motions inside RNAs can be captured and some macroscopic quantities such as bend angle can be measured. Their experiments showed that the bending angle of HIV-1 TAR RNA decreased with increasing NaCl concentration, which was also predicted by recent coarse-grained models.[140,141,144, 145 ] Thirumalai et al. developed and employed an empirical formula for the WLC model to describe the flexibility of the Azoarcus ribozyme at different monovalent and divalent ion concentrations. Additionally, the corresponding persistence lengths were derived by fitting the empirical formula for the WLC model to the measured experimental data.[142]

Rather than examining the local details of non-helical RNA elements, Fulle and Gohlke studied whole RNA molecules directly. They first modeled an RNA structure as a topological network representation using a constraint counting method, and then ran a simulation with a framework rigid optimized dynamics algorithm (FRODA). The root-mean-square fluctuations of all atoms can be determined after simulations, and then the flexibility can be estimated both locally and globally. Quantitative comparisons between FRODA simulations and NMR experiments showed good agreement for some RNAs, including tRNA and pseudoknots.[131]

Furthermore, the flexibility of the ribosomal exit tunnel, a large RNA-protein complex, has been studied using this method.[143] Very recently, a python-based software package named constraint network analysis (CNA) was developed by Gohlke et al.[146] Using the CNA algorithm, one can obtain both global and local properties for the input biomolecules, including RNAs. However, such constraint counting in the CNA algorithm may require more physical optimization and extensive validation.

Since structured RNAs differ from DNA and RNA helices whose flexibility can be well quantified by persistence length, stretching modulus, and twisting modulus, characterizing the flexibility of structured RNAs in a straightforward and quantitative manner remains unclear.

## 5. Conclusion and perspective

As described above, extensive experiments and theoretical modeling have revealed that the flexibility of nucleic acids is tightly coupled to several critical factors. First, the states of structures can dominate the flexibility of nucleic acids, such as the states of the ss chain, ds helices, partially melted helices, and more complicated tertiary folds, corresponding to significantly different flexibilities. Second, the sequences of nucleic acids determine their structures and stabilities, and thus strongly influence their flexibility. Third, temperature can directly determine the state of structures of nucleic acids and thus greatly influence their flexibility. Additionally, solution conditions such as metal ions, which can strongly interact with nucleic acids, significantly affect nucleic acid flexibility, particularly multivalent ions. Recent developments in computation facility and molecular force fields have enabled extensive explorations of the flexibility of nucleic acids at the atomic level. However, despite this great progress in understanding the flexibility of nucleic acids, many important elusive problems must be explored. We will discuss several major challenging issues in the following sections.

*Flexibility of DNA on a short length scale*

A recent AFM experiment showed that spontaneous large-angle bends were many times more prevalent than predicted by the WLC model,[115] suggesting that dsDNA may have much higher flexibility on the short length scale.[147] To examine the mechanism for such higher flexibility on the short length scale, atomic-level MD simulations have been employed for two short dsDNAs, and the calculated apparent persistence length can be ~20 nm on a very short scale (~1−2 bp), and exhibits the oscillation periodically between 20 nm and 100 nm.[148] A correlated WLC model has also been developed to explain the experimental findings, while the microscopic mechanism for such proposed correlation remains unclear.[149] Very recently, high-resolution AFM in solution has been used to analyze the effect on a short length scale, which showed that dsDNA can be well described by the WLC model on a scale beyond 2−3 helical turns.[116,117] However, on a length scale below the threshold of 2−3 helical turns, quantification of the flexibility remains limited by the limitations of AFM. Furthermore, how the local kinking and disruption of hydrogen bond in base pairing affect the flexibility of dsDNA is still unknown. Additionally, the effect of substrate in the AFM experiments on the flexibility of dsDNA must be examined, since a theoretical modeling indicated that $Mg^{2+}$-mediated attraction between DNA and substrate can cause DNA softening.[120] On a short length scale, atomistic MD can become a powerful tool, while reliable force fields in MD are essentially required.

*Flexibility of helices of different conformations*

Two typical conformations of nucleic acids include B-form DNA (B-DNA) and A-form RNA (A-RNA). DsDNA and dsRNA are generally present in B-form and A-form, respectively. B-DNA and A-RNA helices show similar helical structures and some of their elastic properties are qualitatively similar, such as bending persistence length and torsional modulus.[14,15,27] However, B-DNA and A-RNA are significantly different for some elastic properties, such as stretching modulus and stretching-twisting coupling. The stretching modulus of B-DNA is ~3 times higher than that of A-RNA.[14] More surprisingly, single-molecule stretching experiments showed that the stretching-twisting coupling parameter of B-DNA was negative, i.e., the DNA stretched by pulling force (4−8 pN) was accompanied by overwinding of the helix, while that of A-RNA was positive. However, very recently, Manning analyzed existing experimental data and concluded that DNA stretching by environmental change such as the decrease of salt concentration is companied by helix unwinding.[72] Therefore, several unanswered questions remain: (1) Why are B-DNA and A-RNA different in stretching modulus and stretching-twisting coupling? (2) How can we unify the results involving the pulling of B-DNA and the analyses based on experiments involving free B-DNA? (3) Is the flexibility of other conformations of nucleic acid helices such as A-DNA and Z-DNA, also different from the flexibility of B-DNA? Further studies, particularly those on the microscopic level, are required to answer these questions.

*Effect of high salt and multivalent salt*

Numerous experiments have shown that the persistence length of dsDNA decreases and stretching modulus increases with increasing salt concentration such as NaCl. However, the available data regarding salt-dependent *P* can be categorized in two ways: (1) *P* will not continue to decrease after NaCl exceeds 0.1 M;[28,66] (2) $P$ will continue to decrease after NaCl exceeds 0.1 M.[68,69] The classic OSF theory supports the former, while recent coarse-grained simulations and theoretical analysis based on the counterion condensation theory agree with the latter. Therefore, for the salt-dependent flexibility of DNA, several questions remain unanswered: (1) To what extent does salt in the solution influence the flexibility of DNA, and what are the relative fractions of electrostatic and intrinsic contributions to the global flexibility of DNA? (2) Why does $P$ of DNA continue to decrease or become nearly invariant after NaCl exceeds 0.1 M? (3) Why does the stretching modulus of DNA increase at higher salt concentration? (4) Further studies are required to understand the salt dependence of other elastic properties such as torsional modulus and stretching-twisting coupling. A series of experiments in combination with molecular modeling at the atomic level is required to resolve these issues.

The limited experiments on the flexibility of DNA in multivalent salt have demonstrated the dramatic roles of multivalent ions. Multivalent ions can cause an apparent decrease in the persistence length and an apparent increase in the stretching modulus of dsDNA compared with monovalent salt.[28] Such decrease in $P$ has been attributed to multivalent ion-mediated intra-chain attractive force,[82,83,150,151] while the increase in $S$ has not been thoroughly explained. Therefore, to systematically quantify the effect of multivalent ions such as $Mg^{2+}$ and $Co(NH_3)_6^{3+}$ on DNA flexibility is still required, particularly on the local deformation of DNA helix induced by multivalent ions. In addition, the effect of multivalent ions on the flexibility of A-RNA has not been widely examined, while previous experiments have shown that ions can bind to an A-RNA in a very different manner to B-DNA.[152,153] This suggests that the (multivalent) ion effect on A-RNA flexibility may be significantly different from that of B-DNA and thus is highly desirable.

*Flexibility of RNA tertiary folds*

The flexibility of a DNA or RNA helix can be well described by the parameters of the elastic theory of linear polymers or elastic rods. However, RNAs are generally in the folded state of complex native structures beyond a perfect helix, and thus to quantitatively describe the flexibility of non-helix RNAs is beyond the description of the elastic parameters for a helix. The bending angle, persistence length, the distribution of radius of gyration, and the distribution of root of mean square deviation (RMSD) have been used to characterize the flexibility of non-helix RNAs. Nevertheless, there are limitations to these methods. For example, the bending angle only works well for the local bending of a helix and two-way junction. The use of persistence length for structured RNAs (e.g., tRNA) is not feasible, and may only be considered as a relative quantity compared with the denatured ss state or secondary state. The distribution of the radius of gyration can be easily measured, but this value only gives a global description of the flexibility. Since a structured RNA is non-uniform in its flexibility over the entire molecule, the distribution of the radius of gyration may be inadequate for completely describing flexibility. The distribution of RMSD can describe the dynamics of all atoms and can be tracked by experiments and atomic-level modeling, while this description may be only convenient for small RNAs. A combination of these quantities may provide a thorough description of the flexibility of an RNA tertiary structure.

Since RNA tertiary structures are more sensitive to temperature and ionic conditions than helices,[154−158] the flexibilities of RNA tertiary structures are more strongly coupled to temperature and ionic conditions, particularly multivalent ions. Furthermore, dehydrated $Mg^{2+}$ and small molecules such as metabolites can interact specifically with RNAs and alter the flexibility of RNAs to aid their functions.[159−162] Therefore, understanding the flexibility of RNA tertiary structures is not well understood and requires further investigation.

In summary, the results of previous studies have greatly enhanced the understanding of nucleic acid flexibility, but many questions remain and require further comprehensive investigation. In the next decade, we expect more surprising findings regarding the flexibility of nucleic acids as well as more related applications.

## 6. Acknowledgments

We are grateful to Feng-Hua Wang, Yuan-Yan Wu and Ya-Zhou Shi for valuable discussions. This work was supported by the National Key Scientific Program (973)-Nanoscience and Nanotechnology (No. 2011CB933600), the National Science Foundation of China grants (11175132, 11575128 and 11374234), and the Program for New Century Excellent Talents (Grant No. NCET 08-0408).

Table 1. Experimental measurements for the flexibility of single-stranded DNAs/RNAs

| ss nucleic acids | References | Ionic conditions | Thermodynamic quantities |
|---|---|---|---|
| 13k nt ssDNA | Bosco et al[36] | 10−1000 mM $Na^+$; 0.5−10 mM $Mg^{2+}$ | $S, L, P$ |
| 10.5k nt ssDNA | McIntosh et al[38] | 20−3500 mM $K^+$; 20−2000 mM $Na^+$; 0.2−50 mM $Mg^{2+}$; 0.2−50 mM $Ca^{2+}$ | $P$ |
| dT40, rU40 | Chen et al[41] | 0−800 mM $Na^+$; 0−100 mM $Mg^{2+}$ | $R_{ee}, L, P$ |
| dT30 | Meisburger et al[42] | 20 mM $Na^+$; 0−20 mM $Mg^{2+}$ | $R_g, R_{ee}$ |
| dT8−dT100; dA8−dA50 | Sim et al[43] | 12.5−1000 mM $Na^+$ | $R_g, P$ |
| 12−120 nt ss nucleic acids[a] | Wang et al[47] | 1−1000 mM $Na^+$; 0.03−300 mM $Mg^{2+}$; 0.01−100 mM $Co^{3+}$ | $R_{ee}, P$ |
| dT12,dT24,dA12,dA24 | Mills et al[50] | 8−64 mM $Na^+$; 0−8 mM $Mg^{2+}$ | $P$ |
| Poly(U) | Seol et al[51] | 5−500 mM $Na^+$ | $S, P$ |
| 280−5386 nt ssDNA | Tinland et al[52] | 1−100 mM EDTA | $R_g, P$ |

[a] This was a computational study that collected various experimental data for $P$ of ss nucleic acids;

$R_g$: radius of gyration; $R_{ee}$: end-to-end distance; $L$: contour length; $P$: bending persistence length; $S$: stretching modulus.

**Table 2.** Experimental measurements for the flexibility of dsDNAs/dsRNAs

| dsDNAs or dsRNAs | References | Ionic conditions & Temperature | Thermodynamic quantities |
|---|---|---|---|
| 4.2k bp & 8.3k bp dsRNA | Abels et al[7] | moderate salt buffer | $P$ |
| 16, 21, 66 & 89 bp dsDNA | Yuan et al[9] | 500 mM $Na^+$ | $R_g$, $P$ |
| 10, 15, 20, 25, 30, 35 bp dsDNA | Mathew-Fenn et al[10] | 100 mM $Na^+$ | $R_{ee}$, $\sigma_R^2$ |
| 4.2k bp dsRNA & 3.4k bp dsDNA | Lipfert et al[14] | 100 mM & 320 mM $Na^+$ | $P$, $C$, $S$, $D$ |
| 4k bp λDNA & 4k bp dsRNA | Herrero-Galán et al[15] | 0−500 mM $Na^+$ | $P$, $S$ |
| 50k bp λDNA | Baumann et al[28] | 1.86−586 mM $Na^+$; $Mg^{2+}$, $Put^{2+}$, $Spd^{3+}$ & $Co(NH_3)_6^{3+}$ | $P$, $S$ |
| 200 bp λDNA | Geggier et al[31] | TBE buffer; 5−60℃ | $P$ |
| 14.8k bp dsDNA | Bryant et al[58] | 100 mM $Na^+$ | $C$ |
| 50k bp λDNA | Strick et al[62] | 10 mM phosphate buffer | $P$, $C$ |
| T7 DNA | Sobel et al[68] | 5−3000 mM $Na^+$ | $R_{ee}$, $P$ |
| 6954±20 bp dsDNA | Borochov[69] | 7.3−4000 mM $Na^+$ | $R_g$, $P$ |
| dsDNA (125bp−23000 bp) | Mantelli et al[73] | 1 mM $Mg^{2+}$ & 1−100 mM $Na^+$ | $P$ |
| 3888 bp dsDNA | Wang et al[74] | $Na^+$, $K^+$, $Mg^{2+}$, $Spd^{3+}$ | $P$, $S$, $L$ |
| pBR322 dsDNA & Φ6 dsRNA | Lang et al[76] | 48−500 mM $NH_4Cl$ | $L$ |
| 685 bp dsDNA | Driessen et al[91] | 60 mM $K^+$; 100 & 150 mM $Na^+$; 23−52°C | $P$ |
| blunt-ended DNA fragments (41−256 bp) | Porschke[92] | 2.4−110 mM $Na^+$; 0.1 & 10 mM $Mg^{2+}$; 2−20℃ | $P$, $R_H$ |
| 200 bp dsDNA | Geggier et al[99] | TBE buffer | $P$ |
| 2743 bp dsDNA | Wiggins et al[115] | 12 mM $Mg^{2+}$ | Bend angle |

$S$: stretching modulus; $P$: bending persistence length; $C$: torsional persistence length; $L$: contour length; $R_{ee}$: end-to-end distance; $R_g$: radius of gyration; $D$: twist–stretch coupling parameter; $\sigma_R^2$: variance of $R_{ee}$; $R_H$: hydrodynamic radius.

**Table 3.** Different methods employed to probe the flexibilities of structural RNAs

| RNAs | References | Methods | Thermodynamic quantities |
|---|---|---|---|
| tRNA (Asp) | Fulle et al[131] | Constraint counting network & FRODA simulation | RMSD |
| tRNA (Phe) | Roh et al[133] | Quasielastic neutron scattering spectroscopy | $R_g$, $P$ |
| HIV-1 TAR RNA | Zacharias et al[134] | Gel electrophoresis & TEB | Bend angle |
| subsequence of a sRNA (DsrA) | De Almeida Ribeiro E et al[135] | SAXS & NMR & ensemble optimization method | $R_g$ |
| bacterial ribosomal A-site RNA | Fulle et al[136] | Normal mode analysis & MD simulation | Binding free energies, RMSD |
| segments of bulge loops | Zacharias et al[137] | TEB | Bend angle |
| segments of symmetric internal loops | Zacharias et al[138] | TEB | Bend angle |
| HIV-1 TAR RNA | Al-Hashimi et al[139] | NMR & RDC & MD | RMSD |
| HIV-1 TAR RNA | Al-Hashimi et al[140][141] | NMR & RDC | RMSD |
| Azoarcus ribozyme & RNase P | Caliskan et al[142] | SAXS & WLC | $P$ |
| ribosomal exit tunnel | Fulle et al[143] | Constraint counting network & FRODA simulation | RMSD |

$P$: bending persistence length; $R_g$: radius of gyration; RMSD: Root-mean-square deviation.